\documentclass{jps-cp}
\usepackage{txfonts}
\usepackage{xcolor}
\usepackage[normalem]{ulem}
\usepackage{iftex}
\ifpdf
  \makeatletter
  \input{pdftex.def}
  \def\Gin@driver{pdftex.def}
  \makeatother
\fi

\newcommand{\bs}{\boldsymbol}

\title{Magnetotropic Signatures of Fulde-Ferrell Pairing in a Pauli-Limited Superconductor}

\author{
Gulnaz \textsc{Rakhmanova}$^{1}$,
Mikhail \textsc{Lemeshko}$^{1}$,
K. A. \textsc{Modic}$^{1}$,
and Ragheed \textsc{Alhyder}$^{1}$
}

\inst{
$^{1}$Institute of Science and Technology Austria (ISTA),
Am Campus 1, 3400 Klosterneuburg, Austria
}

\email{gulnaz.rakhmanova@ist.ac.at}

\recdate{September 20, 2026}

\abst{
Resonant torsion magnetometry measures the angular curvature of the free energy and is sensitive to magnetic anisotropy. We calculate its response for a clean Pauli-limited superconductor with single-plane-wave Fulde-Ferrell pairing. Field rotation drives a first-order transition from the uniform condensate into the finite-momentum phase and a continuous transition into the normal state. The first-order line produces a sharp softening whose integrated weight is fixed by the magnetization jump, so the dip alone does not identify finite momentum. The distinguishing response is a broad softening inside the Fulde-Ferrell phase, caused by relaxation of the gap and pairing momentum, followed by a finite step at the upper critical field.}

\kword{
Fulde-Ferrell-Larkin-Ovchinnikov state,
finite-momentum pairing,
magnetotropic susceptibility,
resonant torsion magnetometry,
Pauli limiting,
Sr$_2$RuO$_4$
}

\begin{document}
\maketitle

\section{Introduction}

In a conventional spin-singlet superconductor, electrons with opposite momenta and spins form Cooper pairs with zero center-of-mass momentum. A magnetic field polarizes the spin populations through the Zeeman interaction, creating a mismatch between their Fermi surfaces that suppresses uniform pairing at the Pauli limit \cite{Clogston1962,Chandrasekhar1962}. Fulde and Ferrell, and independently Larkin and Ovchinnikov, showed that pairing can survive with a finite center-of-mass momentum, which partly compensates this mismatch \cite{Fulde1964,Larkin1964}. Such phases require clean samples and weak orbital depairing \cite{Matsuda2007}. Sr$_2$RuO$_4$ is a relevant system for this study, as its in-plane upper critical field becomes first-order at low temperatures \cite{Yonezawa2013,Kittaka2014}, torque magnetometry reveals a high-field anomaly \cite{Kikugawa2016}, and NMR spectra show features consistent with spatially modulated spin polarization \cite{Kinjo2022}. Magnetostriction also finds a high-field anomaly, although its relation to the NMR boundary remains unsettled \cite{Kittaka2026,Maeno2024}.

These observations suggest a high-field superconducting phase diagram, but the identification of finite-momentum pairing remains challenging. Resonant torsion magnetometry provides complementary information by converting the angular curvature of the free energy into a shift of a mechanical resonance \cite{Modic2018,Shekhter2023}.
This approach provides high sensitivity for micron-scale samples, where conventional magnetization measurements can be challenging. The magnetotropic response contains both magnetization and spin susceptibility contributions, whose relative importance depends on field orientation. An in-plane field isolates the magnetization term, whereas tilted fields probe both contributions. Here, we calculate the magnetotropic response of a clean Pauli-limited superconductor in which finite-momentum pairing is allowed and examine its evolution with magnetic field and angle. We use the angular dependence to investigate how the response at phase boundaries differs from that within the high-field superconducting state.
\section{Measurement Geometry and Theoretical Model}
\subsection{Measurement Principle}

\begin{figure}[h!]
\centering
\includegraphics[width=0.9\textwidth]{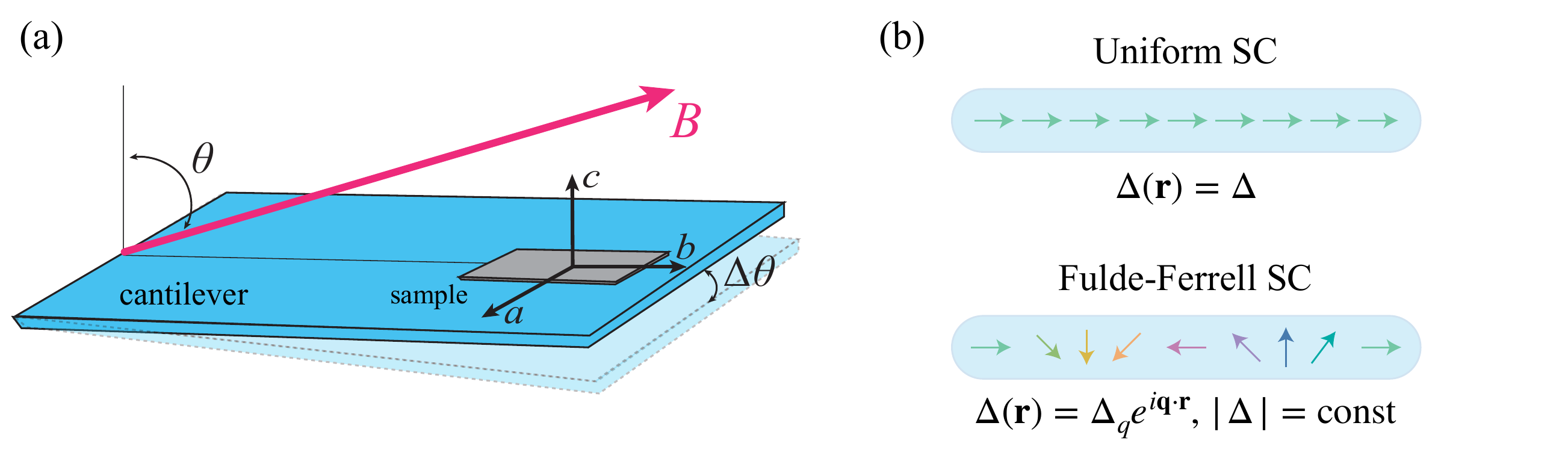}
\caption{Cantilever geometry and superconducting order parameters. (a) A superconducting sample mounted near the free end of a cantilever undergoes small angular oscillations $\delta\theta(t)$ about a field orientation $\theta$. (b) In the uniform state, the order parameter is constant, $\Delta(\mathbf r)=\Delta_0$. In the FF state, $\Delta(\mathbf r)=\Delta_q e^{i\mathbf q\cdot\mathbf r}$, so its magnitude remains constant while its phase winds along $\mathbf q$.}
\label{fig:schematic}
\end{figure}

We consider the cantilever geometry illustrated in
Fig.~\ref{fig:schematic}(a). A superconducting sample of volume $V_s$ is mounted near the free end of a cantilever with mode stiffness $S_n$ and frequency $f_0$. A magnetic field of fixed magnitude makes an angle $\theta$ with the crystalline $c$ axis, while the cantilever motion produces a small angular modulation $\delta\theta(t)$.

 For an equilibrium sample free energy $F_{\rm eq}(\theta)$, the torque is $\tau=-\partial_\theta F_{\rm eq}$ and its linear restoring term is $-k\delta\theta$ with $k=\partial_\theta^2F_{\rm eq}$. While the static torque shifts the mean cantilever position, its angular derivative modifies the stiffness and hence the resonance frequency, $f=f_0\sqrt{1+k/S_n}$. For $|k|\ll S_n$, the relative frequency shift is $\Delta f/f_0\simeq k/(2S_n)$ \cite{Modic2018,Shekhter2023}. To isolate the superconducting contribution, we subtract the normal-state background and define the free-energy density $\mathcal F_{\rm eq}=(F_{\rm eq}-F_{\rm N})/V_s$. The resulting frequency shift is $\Delta f/f_0=V_s\Delta\kappa/(2S_n)$, where $\Delta\kappa=\partial_\theta^2{\cal F}_{\rm eq}$ is the magnetotropic coefficient per unit volume. Positive $\Delta\kappa$ stiffens the mode and negative $\Delta\kappa$ softens it.

\subsection{Single-$q$ superconducting model}

We consider a clean, single-band superconductor with spin-singlet pairing and a cylindrical Fermi surface, characterized by Fermi velocity $v_F$ and density of states $N(0)$ per spin. We set $k_B=\hbar=1$. For a diagonal $g$ tensor, the Zeeman energy is $h(\theta)=b g(\theta)$, where $g(\theta)=(\sin^2\theta+\varepsilon^2\cos^2\theta)^{1/2}$, $b=g_a\mu_BB/2$, and $\varepsilon=g_c/g_a$. The isotropic limit $\varepsilon=1$ gives no angular response, while $\varepsilon=0$ eliminates the Zeeman coupling along $c$. For $0<\varepsilon<1$, the coupling increases as the field is rotated toward the plane and may drive the system through the uniform, FF, and normal phases. Thus, $\varepsilon$ controls the angular positions of the phase boundaries and the magnitude of the magnetotropic response. We use $\varepsilon=0.5$ as an illustrative value. Quantitative comparison with a specific material requires its corresponding $g$-factor anisotropy. Orbital depairing is omitted, so the model is most applicable near orientations where orbital pair breaking is weak.

To describe finite-momentum pairing, we adopt the Fulde--Ferrell (FF) ansatz $\Delta(\bs r)=\Delta e^{i\bs q\cdot\bs r}$, where $\bs q$ is the center-of-mass momentum of the Cooper pairs. This corresponds to pairing between states $(\bs k+\bs q/2,\uparrow)$ and
$(-\bs k+\bs q/2,\downarrow)$. It has a winding phase and constant $|\Delta|$, as shown in Fig.~\ref{fig:schematic}(b).

Near the Fermi surface, $\xi_{\bs k\pm\bs q/2}\simeq\xi_{\bs k}\pm Q\cos\phi$ with $Q=v_Fq/2$. In the corresponding Nambu basis the mean-field Hamiltonian is
\begin{equation}
{\cal H}_{\bs k}=\left(Q\cos\phi-h\right)\tau_0+\xi_{\bs k}\tau_3+\Delta\tau_1,
\label{eq:bdg}
\end{equation}
where the first term describes the residual
mismatch between the Zeeman energy and the
shift associated with finite-momentum pairing. The resulting quasiparticle branches are
$E_\pm=\sqrt{\xi_{\mathbf{k}}^2+\Delta^2}
\pm(h-Q\cos\phi)$. 

Integrating over $\xi_{\mathbf{k}}$ and expressing the pairing interaction in terms of $T_c$ gives the superconducting free-energy density relative to the normal state \cite{Fulde1964,SaintJames1969,Shimahara1994}:
\begin{equation}
\frac{{\cal F}(\Delta,Q,h)}{N(0)}=\Delta^2\ln\frac{T}{T_c}
+2\pi T\sum_{n\ge0}\left\langle\frac{\Delta^2}{\omega_n}
-2\,\mathrm{Re}\left[\sqrt{\tilde\omega_n^2+\Delta^2}-\tilde\omega_n\right]\right\rangle_\phi,
\label{eq:f}
\end{equation}
where $\tilde{\omega}_n=\omega_n+i(h-Q\cos\phi)$,
$\omega_n=\pi T(2n+1)$, and the angular brackets
denote an average over the cylindrical Fermi surface. 

For a cylindrical Fermi surface, the free energy depends only on the magnitude of the pairing momentum, leaving its in-plane direction degenerate.
At each temperature $T$ and Zeeman field $h$, we minimize ${\cal F}$ with respect to $\Delta$ and $Q$ and compare the uniform, FF, and normal states. These minimization conditions yield the gap equation and the zero-current condition, respectively. The global minimum determines the equilibrium values
$\Delta^*$, $Q^*$, and ${\cal F}_{\rm eq}$.
To locate the continuous upper critical line, where
$\Delta$ vanishes, we expand Eq.~(\ref{eq:f}) to
quadratic order in $\Delta$ and minimize its coefficient
over $Q$. Along this line, the optimal $Q$ first becomes nonzero at the Lifshitz tricritical point. The Matsubara sum and angular integration are evaluated numerically, with convergence checked by increasing the cutoff and number of angular points.

\subsection{Magnetotropic response and condensate relaxation}

At fixed $B$ and $T$, rotation changes $h(\theta)$ and the equilibrium values $\Delta_*$ and $Q_*$. The cantilever therefore probes the minimized function ${\cal F}_{\rm eq}[h(\theta)]$, not a curvature at fixed order parameters. The chain rule gives
\begin{equation}
\Delta\kappa(\theta)={\cal F}''(h)\,h_\theta^2+{\cal F}'(h)\,h_{\theta\theta},
\label{eq:kappa}
\end{equation}
where primes are total derivatives with respect to $h$, $h_\theta=b(1-\varepsilon^2)\sin\theta\cos\theta/g$ and $h_{\theta\theta}=\partial_\theta h_\theta$. The first term is the susceptibility contribution, and the second is the magnetization contribution \cite{Shekhter2023}. Their geometric factors vary independently with angle, which allows suitable orientations to emphasize one part of the response. Since ${\cal F}$ is defined relative to the normal-state free energy, its equilibrium derivative satisfies ${\cal F}'=-\delta M$, where $\delta M=M_s-M_N$. Consequently, ${\cal F}''=\chi_N-\chi_s$, with $\chi_N=2N(0)$. For $x=(\Delta,Q)$ and order-parameter Hessian $H_{ij}=\partial^2{\cal F}/\partial x_i\partial x_j$, differentiation of the stationarity conditions gives
\begin{equation}
{\cal F}''
= {\cal F}_{hh}
- {\cal F}_{hx_i} (H^{-1})_{ij}{\cal F}_{x_jh}.
\label{eq:relaxation}
\end{equation}
The first term describes the curvature at fixed order parameters, while the second accounts for their relaxation. Since $H$ is positive definite on a stable branch, the relaxation term lowers the free-energy curvature relative to a calculation with fixed order parameters. It can therefore enhance $\chi_s=\chi_N-{\cal F}''$ above $\chi_N$. Equations (\ref{eq:kappa}) and (\ref{eq:relaxation}) connect the microscopic relaxation of the condensate to the measured frequency shift. 

\section{Phase Diagram and Spin Response}

\begin{figure}[h!]
\centering
\includegraphics[width=0.9\textwidth]{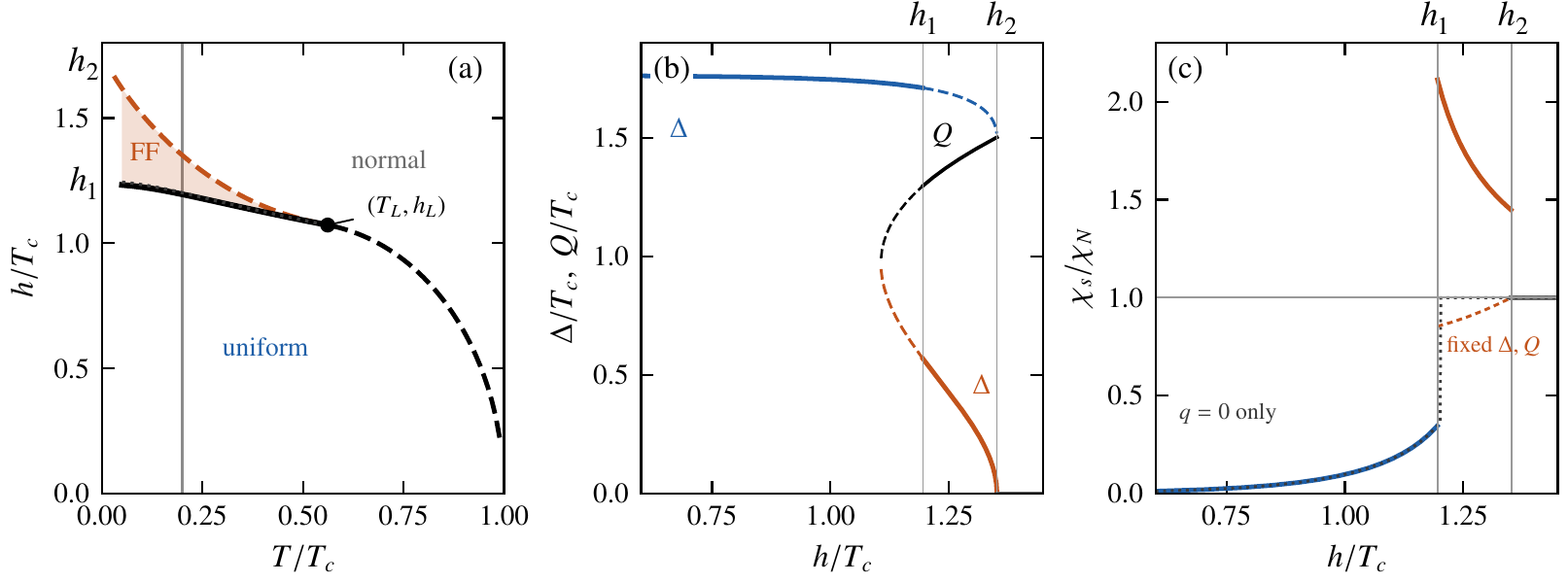}
\caption{Mean-field thermodynamics.
(a) Phase diagram in the $(T,h)$ plane. The boundaries $h_1$ and $h_2\equiv h_{c2}$ mark the first-order uniform-to-FF and continuous FF-to-normal transitions, respectively. The dotted curve is the $q=0$ Pauli boundary, and the vertical line indicates $T=0.2T_c$, used in panels (b) and (c).
(b) Gap $\Delta$ of the uniform (blue) and FF (orange) solutions and pairing momentum $Q=v_Fq/2$ (black) along the $T=0.2T_c$ cut. Solid segments denote the stable equilibrium phases, whereas dashed segments show metastable continuations.
(c) Spin susceptibility $\chi_s=dM/dh$, normalized by the normal-state value $\chi_N$. The solid blue and orange curves show the equilibrium susceptibility in the uniform and FF
phases, including relaxation of $\Delta$ and $Q$. The dashed orange curve shows the FF susceptibility evaluated with $\Delta$ and $Q$ held fixed. The dotted curve is the $q=0$ control, and the horizontal gray line denotes the normal-state value.}
\label{fig:phases}
\end{figure}

Fig.~\ref{fig:phases}(a) shows the mean-field phase diagram. Above the Lifshitz point $(T_L, \,h_L)=(0.561, 1.073) T_c$, the uniform superconducting state undergoes a continuous transition into the normal state \cite{SaintJames1969}. Below this point, the FF phase occupies a finite field interval between the first-order transition at $h_1$ and the continuous transition at $h_{c2}$ \cite{Burkhardt1994,Shimahara1994}. For comparison, the $q=0$ model exhibits a Pauli-limited transition at $h_P$, only slightly above $h_1$ \cite{Sarma1963}. Thus, the position of the first-order transition alone does not establish finite-momentum pairing.

At $T=0.2T_c$, the evolution of the gap $\Delta$ and pairing momentum $Q$  is shown in Fig.~\ref{fig:phases}(b). At $h_1=1.196T_c$, the gap decreases discontinuously from $1.71T_c$ to $0.57T_c$, while $Q$ jumps from zero to $1.30T_c$. With increasing field, the FF gap decreases continuously and vanishes at $h_{c2}=1.352T_c$.  The dashed metastable branches indicate possible hysteresis if switching is delayed.

Fig.~\ref{fig:phases} (c) shows the corresponding spin susceptibility $\chi_s=dM/dh$, normalized to its normal-state value $\chi_N$. In the uniform superconducting state, the susceptibility is suppressed below $\chi_N$ because singlet pairing suppresses spin polarization. On entering the FF phase, the susceptibility (solid orange curve) increases sharply to $\chi_s/\chi_N=2.1$ at $h_1$ and remains enhanced throughout the FF region, reaching $1.45$ just below $h_{c2}$. This curve includes relaxation of both $\Delta$ and $Q$. For comparison, the dashed orange curve shows the response with $\Delta$ and $Q$ held at their initial equilibrium values. The enhancement above $\chi_N$ disappears in this case, demonstrating that it originates from the readjustment of the condensate. The relaxation lowers the gap $\Delta$ and raises the optimal pairing momentum $Q$, allowing a larger portion of the Fermi surface to support spin-polarized quasiparticles. At $h_{c2}$, the susceptibility returns to its normal-state value, producing a finite step in the free-energy curvature, ${\cal F}''(h_{c2}^-)=-0.89\,N(0)$.

Since $\chi_s=dM/dh$, the enhanced susceptibility means that the magnetization increases more rapidly with field within the FF phase than in the normal state. At the first-order boundary $h_1$, the magnetization also exhibits a discontinuous jump $\Delta M=2.06N(0)T_c$, corresponding to approximately $86\%$ of $\chi_N h_1$, so the FF state is already strongly polarized at its lower boundary. Together, the positive magnetization jump and the negative slope of $h_1(T)$ imply through $dh_1/dT=-\Delta S/\Delta M$ that the FF phase has the larger entropy at the transition. These two effects appear differently in the magnetotropic response: the smooth enhancement of $\chi_s$ modifies the regular response within the FF phase, whereas the discontinuous jump $\Delta M$ produces the singular contribution at the first-order boundary discussed below.

\section{Resonator Signal}
Let $\theta_1$ denote the angle at which the first-order transition from the uniform to the FF state occurs, such that $h(\theta_1)=h_1$. At this angle, the equilibrium free energy develops a kink, giving a singular contribution to its angular curvature, $\Delta\kappa(\theta)=W_1\delta(\theta-\theta_1)$, where $W_1=-h_\theta(\theta_1)\Delta M$ and $\Delta M$ is the magnetization jump from the uniform to the FF state. Although its value is concentrated at a single angle, its integrated weight is finite, $\int d\theta\,\Delta\kappa=W_1$. Since the sample torque is $\tau=-V_s\,d{\cal F}_{\rm eq}/d\theta$, the weight corresponds to a finite torque jump $\Delta\tau=-V_sW_1$.

During the measurement, the cantilever oscillates as $\theta(t)=\theta_0+A\cos(\omega t)$. The sample torque, therefore, modifies the effective angular stiffness of the cantilever and shifts its resonance frequency. In the weak-coupling limit, this frequency shift is
\begin{equation}
\left(\frac{\Delta f}{f_0}\right)_A=\frac{V_s}{2S_n}\int_{-A}^{A}dx\,{\cal K}_A(x)\Delta\kappa(\theta_0+x),
\qquad {\cal K}_A(x)=\frac{2\sqrt{A^2-x^2}}{\pi A^2}.
\label{eq:finite_amplitude}
\end{equation}
Here ${\cal K}_A(x)$ is the normalized finite-amplitude kernel. Convolution with this kernel transforms the ideal delta-function singularity into the finite line shape $W_1{\cal K}_A(\theta_1-\theta_0)$ with width $2A$ and extremal curvature $2W_1/(\pi A)$. Its integrated weight remains $W_1$, while its peak magnitude scales as $1/A$.

\begin{figure}[h!]
\centering
\includegraphics[width=\textwidth]{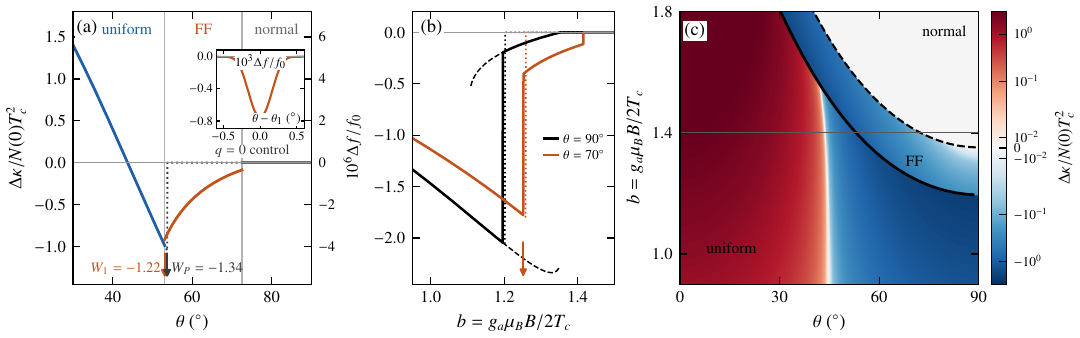}
\caption{Magnetotropic response at $T=0.2T_c$ and $\varepsilon=0.5$.
(a) Angle sweep at fixed $b=1.40T_c$. Blue, orange, and gray curves show the regular response in the uniform, FF, and normal phases, respectively. The dotted curve is the $q=0$ control. Arrows show the integrated weights of the singular first-order contributions. The inset shows the broadened frequency shift line shape near $\theta_1$ for oscillation amplitude $A=0.2^\circ$ and angular spread $0.1^\circ$.
(b) Field sweeps at $\theta=90^\circ$ (black) and $\theta=70^\circ$ (orange). Solid curves show the equilibrium finite-$q$ response, dotted curves the $q=0$ control, and dashed curves metastable continuations.
(c) Regular response in the field-angle plane. Red and blue denote positive and negative $\Delta\kappa$, respectively. The solid and dashed black lines mark the uniform-to-FF and FF-to-normal boundaries, and the horizontal line corresponds to the angle sweep in panel (a).}
\label{fig:sweeps}
\end{figure}

Fig.~\ref{fig:sweeps} summarizes the predicted magnetotropic response at $T=0.2T_c$ and $\varepsilon=0.5$. Fig.~\ref{fig:sweeps}(a) shows an angle sweep at fixed $b=1.40T_c$. At the first-order uniform-to-FF transition, $\theta_1=53.1^\circ$, the magnetization jump produces a singular contribution with integrated weight $W_1=-1.22\,N(0)T_c^2$, indicated by the arrow. Finite cantilever amplitude and angular spread broaden this singularity into the dip shown in the inset, with peak frequency shift $\Delta f/f_0\simeq-7.5\times10^{-4}$. Its weight is close to the $q=0$ value, $W_P=-1.34\,N(0)T_c^2$, so the first-order dip alone does not distinguish the FF phase. The clearer difference is the negative response that persists across the approximately $20^\circ$ FF interval and ends at the continuous transition $\theta_2=72.5^\circ$.

Fig.~\ref{fig:sweeps}(b) compares field sweeps at $\theta=90^\circ$ and $70^\circ$. For an in-plane field, $h_\theta=0$, so the response depends only on the magnetization difference, $\Delta\kappa=(1-\varepsilon^2)\,b\,\delta M$. At $70^\circ$, both the magnetization and susceptibility terms are present. Immediately below $h_{c2}$, the response is $-0.11\,N(0)T_c^2$, which jumps to zero upon entering the normal state. Comparing these orientations, therefore, provides complementary information about $\delta M$ and $\chi_s$. Fig.~\ref{fig:sweeps}(c) combines these results in the field-angle plane. The FF phase appears as a negative-response band between the first-order and continuous phase boundaries. This band disappears when $q$ is constrained to zero, leaving the geometric sign change near $44^\circ$. Deviations from the predicted dependence on $h(\theta)$ would indicate physics beyond the Zeeman-only model, potentially including orbital effects.

For representative Sr$_2$RuO$_4$ parameters \cite{Mackenzie2003}, the predicted frequency-shift scale is approximately $4\times10^{-6}$ per plotted unit, above the demonstrated sensitivity of $6\times10^{-9}$ \cite{Modic2018}. The actual signal will depend on background subtraction and on how orbital depairing modifies the FF region.

\section{Discussion}

The first-order dip is fixed by the magnetization jump and is not unique to finite momentum. Its integrated weight is the robust quantity, whereas its height depends on the cantilever amplitude and the angular spread of the sample. The nearly equal weights of the FF and $q=0$ transitions show that this dip locates a thermodynamic boundary but cannot by itself identify the state on its high-field side.

A more selective prediction is the broad softening within the FF phase, followed by a finite step at the upper critical field. An experimental analysis can separate these features by repeating the sweep at several oscillation amplitudes. The singular peak should scale approximately as $1/A$ while retaining its area, whereas the regular FF background and the upper-boundary step should remain unchanged. Comparing field sweeps at two orientations then separates the magnetization and susceptibility terms in Eq.~(\ref{eq:kappa}). Reversing the sweep provides an independent check for metastability, while a simultaneous dissipation signal would reveal delayed switching.

Quantitative modeling of Sr$_2$RuO$_4$ requires the competing Larkin--Ovchinnikov state, orbital depairing, multiband structure and spin-orbit-entangled pseudospins \cite{Matsuda2007, Kikugawa2016, Mackenzie2003, Maeno2024}. These effects can shift the phase boundaries, modify the response, or restrict the stability of finite-momentum pairing. They do not alter the main distinction between a localized transition dip and a softening. Magnetotropic measurements can therefore test the thermodynamics inside the proposed high-field state rather than only locate its boundary.




\begin{thebibliography}{99}

\bibitem{Clogston1962}
A. M. Clogston, Phys. Rev. Lett. \textbf{9}, 266 (1962).

\bibitem{Chandrasekhar1962}
B. S. Chandrasekhar, Appl. Phys. Lett. \textbf{1}, 7 (1962).

\bibitem{Fulde1964}
P. Fulde and R. A. Ferrell, Phys. Rev. \textbf{135}, A550 (1964).

\bibitem{Larkin1964}
A. I. Larkin and Yu. N. Ovchinnikov, Zh. Eksp. Teor. Fiz. \textbf{47}, 1136 (1964) [Sov. Phys. JETP \textbf{20}, 762 (1965)].

\bibitem{Matsuda2007}
Y. Matsuda and H. Shimahara, J. Phys. Soc. Jpn. \textbf{76}, 051005 (2007).

\bibitem{Yonezawa2013}
S. Yonezawa, T. Kajikawa, and Y. Maeno, Phys. Rev. Lett. \textbf{110}, 077003 (2013).

\bibitem{Kittaka2014}
S. Kittaka \textit{et al.}, Phys. Rev. B \textbf{90}, 220502(R) (2014).

\bibitem{Kikugawa2016}
N. Kikugawa \textit{et al.}, Phys. Rev. B \textbf{93}, 184513 (2016).

\bibitem{Kinjo2022}
K. Kinjo \textit{et al.}, Science \textbf{376}, 397 (2022).

\bibitem{Kittaka2026}
S. Kittaka \textit{et al.}, J. Phys. Soc. Jpn. \textbf{95}, 014704 (2026).

\bibitem{Maeno2024}
Y. Maeno, S. Yonezawa, and A. Ramires, J. Phys. Soc. Jpn. \textbf{93}, 062001 (2024).

\bibitem{Modic2018}
K. A. Modic \textit{et al.}, Nat. Commun. \textbf{9}, 3975 (2018).

\bibitem{Shekhter2023}
A. Shekhter \textit{et al.}, Phys. Rev. B \textbf{108}, 035111 (2023).

\bibitem{SaintJames1969}
D. Saint-James, G. Sarma, and E. J. Thomas, \textit{Type II Superconductivity} (Pergamon, Oxford, 1969).

\bibitem{Burkhardt1994}
H. Burkhardt and D. Rainer, Ann. Phys. (Leipzig) \textbf{3}, 181 (1994).

\bibitem{Shimahara1994}
H. Shimahara, Phys. Rev. B \textbf{50}, 12760 (1994).

\bibitem{Sarma1963}
G. Sarma, J. Phys. Chem. Solids \textbf{24}, 1029 (1963).

\bibitem{Mackenzie2003}
A. P. Mackenzie and Y. Maeno, Rev. Mod. Phys. \textbf{75}, 657 (2003).

\end{thebibliography}
\end{document}